\documentclass[preprint]{aastex}
\begin{document}

\title{Structure and Composition of Molecular Clouds with CN Zeeman Detections I: W3OH} 
\author{Nicholas S. Hakobian and Richard M. Crutcher} 
\affil{University of Illinois at Urbana-Champaign} 
\affil{1002 W. Green Street, Urbana IL} 
\email{nhakobi2@astro.illinois.edu, crutcher@illinois.edu}

\begin{abstract}
We have carried out a multi-species study of a region which has had
previous measurements of strong magnetic fields through the CN Zeeman
effect in order to to explore the relationship between CN and
N$_2$H$^+$, both of which have evidence that they remain in the gas
phase at densities of 10$^5$ - 10$^6$ cm$^{-3}$.  To achieve this we
map the 1 arcmin$^2$ region around the UCHII region of W3(OH) using
the Combined Array for Millimeter-wave Astronomy
(CARMA). Approximately 105 hours of data were collected in multiple
array configurations to produce maps with an effective resolution of
$\sim$ 2.5\arcsec at high signal-to-noise in CN, C$^{18}$O, HCN,
HCO$^+$, N$_2$H$^+$, and two continuum bands (91.2 GHz and 112
GHz). These data allow us to compare tracer molecules associated with
both low and high density regions to infer gas properties.  We
determine that CARMA resolves out approximately 35\% of the CN
emission around W3(OH) when compared with spectra obtained from the
IRAM-30 meter telescope. The presence of strong absorption lines
towards the continuum source in three of the molecular transitions
infers the presence of a cold, dark, optically thick region in front
of the continuum source. In addition, the presence of high-velocity
emission lines near the continuum source shows the presence of hot
clumpy emission behind the continuum source.  These data determine
that future high-resolution interferometric CN Zeeman measurements
which cannot currently be performed (due to technical limitations of
current telescopes) are feasible. We confirm that CN is indeed a good
tracer for high density regions; with certain objects such as W3(OH)
it appears to be a more accurate tracer than N$_2$H$^+$.
\end{abstract}

\section{Introduction}

The measurement of magnetic fields harbor several difficulties due to
the fact that they are a vector quantity. In order to measure the
absolute magnitude and direction of the magnetic field, it is
necessary to sample the two components in the plane of the sky and the
line-of-sight component. While dust polarization can measure the
direction of a magnetic field in the plane of the sky, the
line-of-sight component intensity can be directly measured. This is
accomplished through the detection of the normal Zeeman effect which
occurs when an atom or molecule is in the presence of a magnetic
field. Most transitions have a Zeeman splitting factor that is too
small to measure unless in dense, spatially-compact environments with
exceedingly large magnetic field strengths. However, HI, OH, and CN
have splitting factors that are large enough to measure the weaker
magnetic fields that are expected around sites of active star
formation in molecular clouds. The CN N = 1-0 transition has seven
strong hyperfine components (Table \ref{cntable}) that have a large
distribution of Zeeman splitting factors. The two strongest hyperfine
components, with relative strengths of 27 and 10 respectively, also
have large splitting factors.

\begin{deluxetable}{lcccc}
\tablecaption{CN Hyperfine Components\label{cntable}}
\tablehead{\colhead{Molecule} & \colhead{Transition} & \colhead{$\nu$ (MHz)} 
            & \colhead{Z (Hz/$\mu$G)\tablenotemark{\emph{a}}} 
            & \colhead{RI\tablenotemark{\emph{b}}}}
\tablewidth{0pt}
\startdata
CN & N=$1-0$ J=$\frac{1}{2}-\frac{1}{2}$ F=$\frac{1}{2}-\frac{1}{2}$ 
   & 113123.337 & --\tablenotemark{\emph{c}} & 1 \\
CN & N=$1-0$ J=$\frac{1}{2}-\frac{1}{2}$ F=$\frac{1}{2}-\frac{3}{2}$ 
   & 113144.192 & 2.18 & 8 \\
CN & N=$1-0$ J=$\frac{1}{2}-\frac{1}{2}$ F=$\frac{3}{2}-\frac{1}{2}$ 
   & 113170.528 & -0.31 & 8 \\
CN & N=$1-0$ J=$\frac{1}{2}-\frac{1}{2}$ F=$\frac{3}{2}-\frac{3}{2}$ 
   & 113191.317 & 0.62 & 10 \\
CN & N=$1-0$ J=$\frac{3}{2}-\frac{1}{2}$ F=$\frac{3}{2}-\frac{1}{2}$ 
   & 113488.140 & 2.18 & 10 \\
CN & N=$1-0$ J=$\frac{3}{2}-\frac{1}{2}$ F=$\frac{5}{2}-\frac{3}{2}$ 
   & 113490.982 & 0.56 & 27 \\
CN & N=$1-0$ J=$\frac{3}{2}-\frac{1}{2}$ F=$\frac{1}{2}-\frac{1}{2}$ 
   & 113499.639 & 0.62 & 8 \\
CN & N=$1-0$ J=$\frac{3}{2}-\frac{1}{2}$ F=$\frac{3}{2}-\frac{3}{2}$ 
   & 113508.944 & 1.62 & 8 \\
CN & N=$1-0$ J=$\frac{3}{2}-\frac{1}{2}$ F=$\frac{1}{2}-\frac{3}{2}$ 
   & 113520.414 & --\tablenotemark{\emph{c}} & 1 \\
\enddata
\tablenotetext{\emph{a}}{Z is the Zeeman Splitting Factor for each hyperfine 
  component of CN.}
\tablenotetext{\emph{b}}{RI is the relative intensity of each component.}
\tablenotetext{\emph{c}}{The Z's for these hyperfine components were not calculated due to their low RI.}
\tablerefs{Falgarone et al. 2008}
\end{deluxetable}

\citet{falgarone08} surveyed 14 molecular cloud cores and measured the
line-of-sight magnetic field strength using the CN Zeeman effect. The
strongest magnetic field detected was $1.10\pm 0.33$ mG towards the
Ultra-Compact HII (UCHII) region of W3(OH) \citep{hoare04}, a high
mass star formation site in the W3 molecular cloud complex. For these
measurements, \citet{falgarone08} used the IRAM 30-meter
telescope. Due to the relatively large beam size of the IRAM-30 meter
telescope (23\arcsec\ or 46,000 AU in diameter at the 2 kpc distance
\citep{hachisuka06} of W3(OH)), only very limited mapping to find the
emission peak was performed. From these data, it is not possible to
determine the structure of the magnetic field or which material the
field is associated with.

W3(OH) contains a high-mass O star and has been extensively studied
over the years since the discovery of several OH maser sites within it
\citep{raimond69}. A star forming region ~6\arcsec\ east of W3(OH) was
later discovered by \citet{turner84} in HCN. This smaller region was
also studied for maser activity and found to contain several H$_2$O
masers \citep{wynnwilliams72}. While this second source was initially
not observed in the continuum, with the development of more sensitive
instruments, it was eventually detected in dust continuum studies
\citep{wilner95,wyrowski97}. Within the last few years, further high
resolution studies with instruments such as BIMA have led to the
detection of multiple dense cores believed to be sites of active star
formation within the Turner-Welch Object \citep{chen06}.

In order to determine the spatial distribution of the magnetic field,
an understanding of the CN gas is needed. There is some question,
however, as to the exact physical conditions that CN
traces. \citet{hilyblant08} used the IRAM 30-m telescope to conclude
that CN in prestellar cores stays in the gas phase at densities close
to $10^6$ cm$^{-3}$ and can serve as a kinematic tracer of high
density gas. They compared their CN maps with N$_2$H$^+$, another high
density tracer, and found the results to be in good
agreement. Therefore, it is expected that the measured magnetic field
strength derived from CN measurements would be in gas that is
associated with high density regions of active star
formation. However, before detailed conclusions can be drawn from
these magnetic field measurements, a high angular resolution study of
the CN gas distribution is required. An ideal instrument for this is
an interferometer such as CARMA, as its high spatial resolution and
ability to sample baselines as short as 6m allow us to sample material
both spatially compact and widespread.

\section{CARMA Observations}

With CARMA \citep{bock06} it is possible to achieve a resolution of
approximately 2.5$\arcsec$ by combining maps generated with the C, D,
and E arrays. By using three different array configurations, we have
the ability to probe small scale structure while retaining the ability
to image large scale features that would otherwise be resolved out in
the C and D arrays. CARMA currently does not have polarization
capability and therefore cannot perform magnetic field measurements;
however, by mapping these regions at high resolution we can gain
crucial information as to the structure of the regions that are being
sampled by the IRAM observations. To assist with this, we use the
advanced features of CARMA's correlator to sample several spectral
lines simultaneously, so we can compare the structure of the CN
emission with emission of other well studied tracer molecules.  We
will also gain information on which regions are possible to map in the
Zeeman effect at high resolution and for long integration times with
an instrument capable of making dual-polarization measurements, when
such an interferometer array becomes available.

The CARMA observations began in Spring 2007 and were completed in Fall
of 2009. About 66 hours of observing time was used to map a 1
arcmin$^2$ region around W3(OH). This time was used to observe the
following molecular tracers: CN, C$^{18}$O, N$_2$H$^+$, HCN, and
HCO$^+$ (Table \ref{transitionstable}). Due to the design of the CARMA
correlator and the relatively large frequency range within the 3-mm
band that these tracers span, the measurements had to be performed in
two separate tracks: one for CN and C$^{18}$O, and a second for
N$_2$H$^+$, HCN, and HCO$^+$. Approximately 50 hours (Table
\ref{w3ohtime}) was spent for the CN band and 16.2 hours was spent for
the N$_2$H$^+$ band (not counting additional time spent in B array,
discussed below). In order for us to reach our signal to noise goal,
the tracks containing N$_2$H$^+$ required much less observing
time. Table \ref{w3ohtime} shows the observing array details. Our
pointing center is located on the continuum source, an ultra-compact
HII region, located at: 02:27:03.7 RA, +61:52:25 DEC (J2000). This is
slightly offset from the IRAM pointing position.

\begin{deluxetable}{lcr}
\tablecaption{Observed Transitions\label{transitionstable}}
\tablehead{\colhead{Molecule} & \colhead{Transition} & \colhead{$\nu$ (MHz)}}
\tablewidth{0pt}
\startdata
HCN & N=$1-0$ F=$1-1$ & 88630.415 \\
HCN & N=$1-0$ F=$2-1$ & 88631.847 \\
HCN & N=$1-0$ F=$0-1$ & 88633.936 \\
HCO$^+$ & N=$1-0$ & 89188.526 \\
N$_2$H$^+$ & N=$1-0$ F$_1$=$1-1$ F=$0-1$ & 93171.621 \\
N$_2$H$^+$ & N=$1-0$ F$_1$=$1-1$ F=$2-2$ & 93171.917 \\
N$_2$H$^+$ & N=$1-0$ F$_1$=$1-1$ F=$1-0$ & 93172.053 \\
N$_2$H$^+$ & N=$1-0$ F$_1$=$2-1$ F=$2-1$ & 93173.480 \\
N$_2$H$^+$ & N=$1-0$ F$_1$=$2-1$ F=$3-2$ & 93173.777 \\
N$_2$H$^+$ & N=$1-0$ F$_1$=$2-1$ F=$1-1$ & 93173.967 \\
N$_2$H$^+$ & N=$1-0$ F$_1$=$0-1$ F=$1-2$ & 93176.265 \\
C$^{18}$O & N=$1-0$ & 109782.176 \\
CN & N=$1-0$ J=$\frac{3}{2}-\frac{1}{2}$ F=$\frac{3}{2}-\frac{1}{2}$ & 113488.140 \\
CN & N=$1-0$ J=$\frac{3}{2}-\frac{1}{2}$ F=$\frac{5}{2}-\frac{3}{2}$ & 113490.982 \\
\enddata
\end{deluxetable}

\begin{deluxetable}{clll}
\tablecaption{W3OH Observation Lengths\label{w3ohtime}}
\tablehead{\colhead{Array Config} & \colhead{CN, C$^{18}$O} & \colhead{HCN, HCO$^+$, N$_2$H$^+$} & \colhead{Resolution}}
\tablewidth{0pt}
\startdata
C & 28.0h & 7.4h & 1.5\arcsec \\
D & 8.1h  & 4.3h & 5.0\arcsec \\
E & 11.0h & 4.5h & 10.0\arcsec \\
B & 30h & 12h & $\sim 0.7$\arcsec \\
\enddata
\end{deluxetable}

We produced maps for 3-mm lines of CN, C$^{18}$O, N$_2$H$^+$, HCN, and
HCO$^+$ at a resolution of approximately 2.5$\arcsec$. The composite
CARMA primary beam at half power at this frequency is slightly more
than 60$\arcsec$ in diameter, reflected in our maps, which are
64$\arcsec$ on a side. Since baselines as short as 6m are included,
the maps are sensitive to structure smaller than about 90$\arcsec$, so
that the spatial dynamic range is almost 40:1. At the distance of
W3(OH) (2 kpc), this means the maps are about 0.62 pc across, and the
resolution is about 0.024 pc or 5,000 AU. For all of the spectral line
observations, we used the 8 MHz spectral line mode, which for CN
(N$_2$H$^+$) provides a velocity coverage of $\sim$21 km s$^{-1}$
($\sim$25.8 km s$^{-1}$) and a resolution of 0.335 km s$^{-1}$ (0.415
km s$^{-1}$). The other transitions have similar coverages and
resolutions which vary depending on their rest frequency. The large
velocity coverage is necessary to simultaneously image several
hyperfine components while still maintaining a high spectral
resolution. Our data also contains two 500 MHz continuum bands in each
track giving us four separate continuum windows between 88 GHz and 113
GHz. These continuum maps, while having a lower spatial resolution
than some previously published maps of W3(OH) \citep{chen06}, have
a significantly higher signal-to-noise ratio.

MIRIAD was used for data reduction, specifically with the
modifications for use with CARMA. Due to the nature of CARMA, with 3
distinct primary beams, the data were reduced as a mosaic data set
with a common pointing center.  Passband calibration was performed on
all tracks by utilizing the CARMA system noise source. Even though
suitable passband calibrators were observed for every track, it was
decided that better passband solutions could be obtained from the
noise source.  Flux calibration was performed on each individual track
with a 15 minute observation of MWC349, Mars, or Uranus, whichever was
best suitable at observation time as the intrinsic flux of these
sources is very well modeled. Since our data were obtained over such a
long period of time, the flux of our primary phase calibrator varied
significantly. We determine the flux of the phase calibrator through a
bootstrapping process by which our flux calibrator measurements are
used to calibrate the flux scale for the point source phase
calibrator. It is estimated that there is an inherent 20\% error with
this flux calibration technique. Phase calibration was performed
primarily on 0359+509 with a secondary calibrator 0102+584, used if
the primary was not visible for a significant portion of the
track. The primary phase calibrator had an averaged bootstrapped flux
of 6.16 Jy over 10 tracks obtained over a period of 20 months. The
flux steadily increased from 3.9 Jy to 10.7 Jy over this time. In
order to corroborate the flux measurements we compared our data with
that obtained by the CARMA flux calibration commissioning task. This
independent flux measurement of 0359+509 is 9.83 Jy (from Fall 2009),
consistent with our measurement of 10.7 Jy from the same time period,
assuming the 20\% uncertainty mentioned above. Gain and phase
calibration was performed with the \emph{gfiddle} routine which fits
an n-th order polynomial to the phase measurements on the phase
calibrator. This method was chosen in contrast to the more widely used
\emph{selfcal} technique (on the calibrator) due to the weak signal
strength of some of our spectral lines. However the benefits of this
technique over the selfcal technique have not been shown; overall this
technique may not have any net advantage. It may be possible to gain
an increased signal to noise by performing self-calibration on W3OH
itself, as the continuum source in the center of the map remains
unresolved in all of the array configurations and is bright enough to
perform phase calibration in the spectral channels. However, the
absolute position of the phase center of the map is not retained and
flux calibration is not guaranteed to succeed. These details are
described in detail below in the description of our B-array data.

Since CARMA is composed of two types of telescopes of differing sizes,
we have to handle the fact that we have multiple primary beam types
associated with our data. CARMA has both 10-meter and 6-meter dishes
which form 3 effective primary beams with the same phase center: one
for the pairing of 10-meter to 10-meter dish, one for 6-meter to
6-meter, and a third of 10-meter to 6-meter. At the CN line frequency
the 6-meter primary beam half power beam width (HPBW) is
$\sim$111\arcsec, the 10-meter HPBW is $\sim$66\arcsec (at the
N$_2$H$^+$ frequency of 93 GHz, these HPBWs are 135\arcsec\ and
81\arcsec, respectively), while the 6m-10m primary beam size is in
between that of the other two. We conservatively constrained our maps
to $\sim$1 arcmin$^2$ in order to easily compare the other maps with
the CN data. The restoring beam size is calculated by fitting the
dirty beam with a Gaussian beam. This is the effective resolution of a
map, and is the value quoted in all of our figures.

Image conversion from the UV-dataset into the spatial domain was
performed by the MIRIAD routine \emph{invert}. System temperature
weighting was used to properly downweight data taken at low elevation
or in the unlikely event that data were improperly flagged. The CARMA
control system has an intricate flagging mechanism that operates if
one of many error conditions are met including: telescope tracking
errors, pointing errors, receiver problems (dewar temperature, LO
frequency, etc), system temperature, and other computer or correlator
errors that could hinder data integrity. All data were additionally
inspected for extreme system temperatures, unphysical antenna gains,
time regions without converging phase solutions, and poor weather
conditions that could influence the quality of the data. In a few
cases (usually in older tracks before some automated flagging
conditions were introduced) data that contained systematic errors were
manually flagged. This additional flagging increased the
signal-to-noise of these few tracks by a factor of 2-3.  In the
\emph{invert} step, we weighted our UV data using a Briggs visibility
weighting robustness parameter of 1 \citep{briggs99}, which
provides a slightly increased signal-to-noise ratio over uniform
weighting at the expense of a larger synthesized beam size. The
relatively weak CN emission prompted the weighting of data in this
manner. Since we are detecting low-level emission we did not want
large sidelobes to obscure the detections. If we changed the weighting
more towards Natural weighting, we could have reduced the noise level
further; however, the beam size would be significantly larger and we
would be affected by significant sidelobes. Cleaning was performed
with the MIRIAD task \emph{mossdi}, the mosaic version of the standard
Steer clean routine. We used a multi-step cleaning technique in order
to prevent under-cleaning the source. A small number of clean
iterations was performed in order to remove large scale beam patterns
while leaving much of the low intensity source features
uncleaned. From this initial clean, we ran the MIRIAD task
\emph{restor} which used the clean data to produce a map without beam
effects. We used the average off-line noise level in this map to limit
how much flux was cleaned from a second iteration of \emph{mossdi} and
to prevent overcleaning. This second stage \emph{mossdi} iteration was
followed by another invert step to produce a final, deep cleaned
map. In theory, this process can be repeated ad infinitum, however, if
the noise level calculated in the first step is accurate, and the
maximum number of cleaning iterations is large enough such that we are
guaranteed to clean down to the noise level, this two step process is
adequate.

\section{Results}
\subsection{CN}

\begin{figure}
\begin{center}
\includegraphics[height=\linewidth, angle=-90]{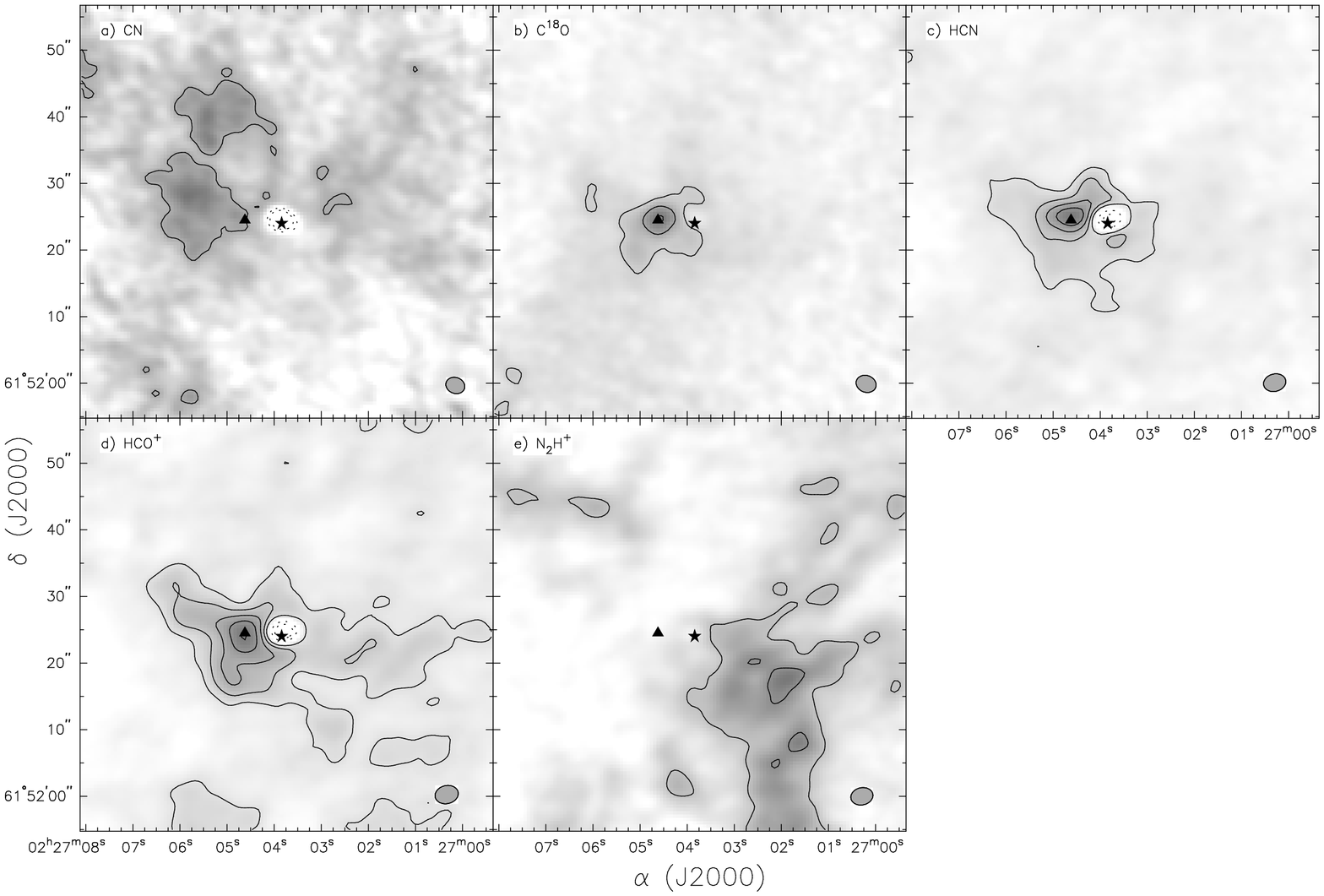}
\end{center}
\caption{CARMA integrated line maps.  \emph{(a)} CN, $2.9\arcsec$ x
  $2.4\arcsec$, offline RMS = 0.03 Jy beam$^{-1}$, Peak intensity =
  0.29 Jy beam$^{-1}$, Peak SNR = 9.6.  \emph{(b)} C$^{18}$O, $3.1\arcsec$
  x $2.5\arcsec$, offline RMS = 0.015 Jy beam$^{-1}$, Peak intensity =
  0.52 Jy beam$^{-1}$, Peak SNR = 34.6.  \emph{(c)} HCN, $3.4\arcsec$ x
  $2.6\arcsec$, offline RMS = 0.021 Jy beam$^{-1}$, Peak intensity =
  0.84 Jy beam$^{-1}$, Peak SNR = 40.  \emph{(d)} HCO$^+$,
  $3.5\arcsec$ x $2.7\arcsec$, offline RMS = 0.037 Jy beam$^{-1}$, Peak
  intensity = 0.86 Jy beam$^{-1}$, Peak SNR = 23.2.  \emph{(e)}
  N$_2$H$^+$, $3.4\arcsec$ x $2.6\arcsec$, offline RMS = 0.023 Jy
  beam$^{-1}$, Peak intensity = 0.41 Jy beam$^{-1}$, Peak SNR =
  17.8. The contour levels are 3, 6.5, 10, 13.5, 17 times 0.05 Jy
  beam$^{-1}$. The $\bigstar$ and $\blacktriangle$ represent the peak
  continuum positions of W3(OH) and the Turner-Welch object,
  respectively.
\label{w3ohmap}}
\end{figure}

\begin{figure}
\begin{center}
\includegraphics[height=\linewidth, angle=-90]{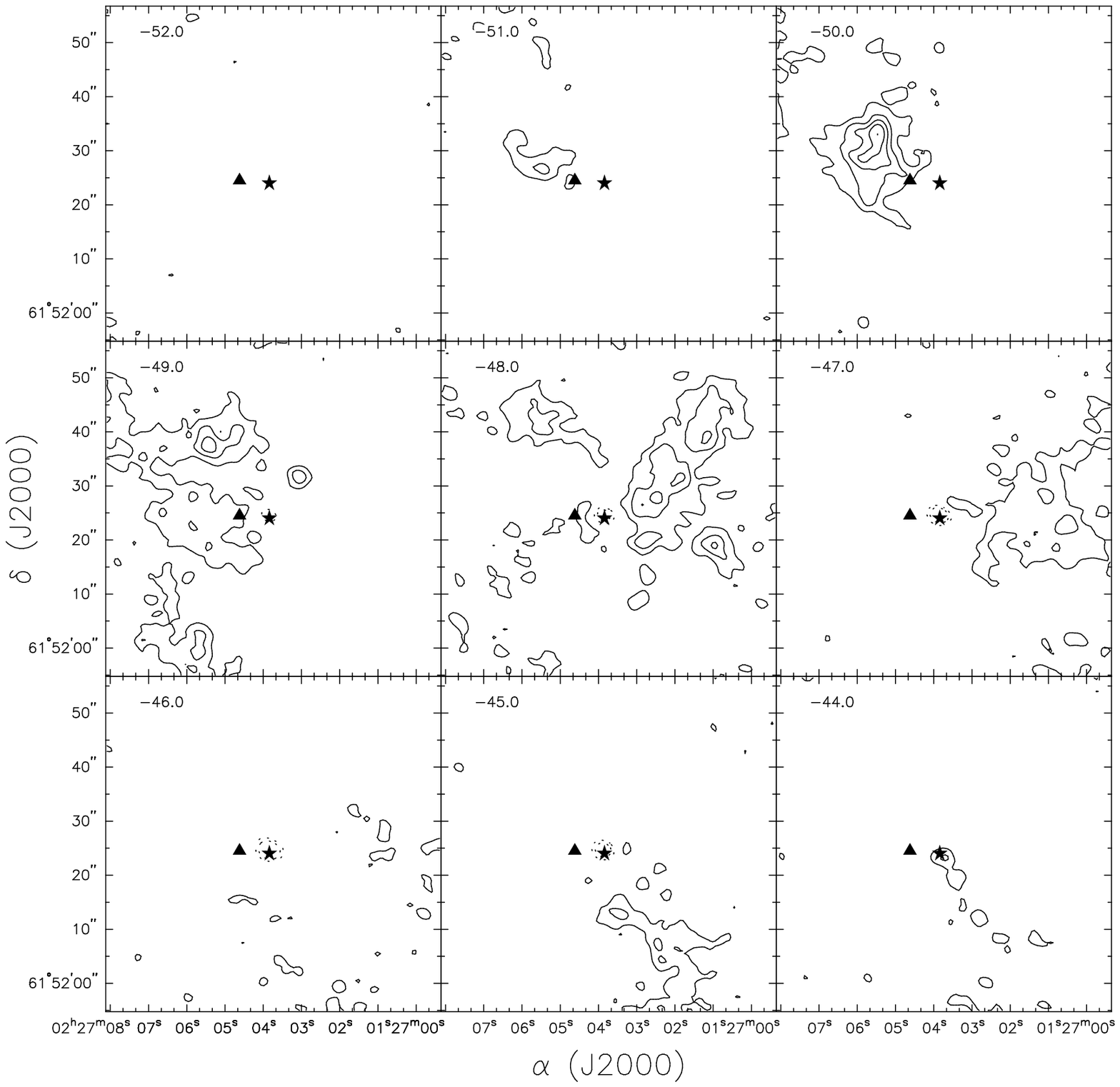}
\end{center}
\caption{CN Channel maps, rebinned to 1 km s$^{-1}$ velocity
  increments. The velocity scale is centered on the strongest CN
  hyperfine component at 113.490 GHz. The spacial resolution of this
  map is 2.9 x 2.4 arcsec. The countour levels are -80, -40, 20, 40,
  60, 80, 100 times 0.00989 Jy beam$^{-1}$. The $\bigstar$ and
  $\blacktriangle$ represent the peak continuum positions of W3(OH)
  and the Turner-Welch object, respectively.
\label{cnchannel}}
\end{figure}

Figure \ref{w3ohmap} shows integrated line maps of the five
species, while Figure \ref{cnchannel} shows channel maps. The continuum
source was removed from the spectral line maps by averaging several
off-line channels together and subtracting that from each individual
channel. Towards the continuum source, we see three of the species
strongly in absorption. Besides the Turner-Welch object and the
continuum source, there are two other regions that are notable in
CN. The western side of the map contains very diffuse CN emission,
while the eastern side of the map shows a compact and complex CN
emission source. In order to compare the CARMA CN data with the IRAM
CN data \citep{falgarone08}, we also generated a map where the
CARMA channel data were convolved with a 23$\arcsec$ Gaussian beam,
effectively smoothing the CARMA data to match the IRAM resolution (see
\S 4.2). By comparing these maps, there is evidence that both spectra
are dominated by diffuse emission.

\begin{figure}
\begin{center}
\includegraphics[height=\linewidth, angle=-90]{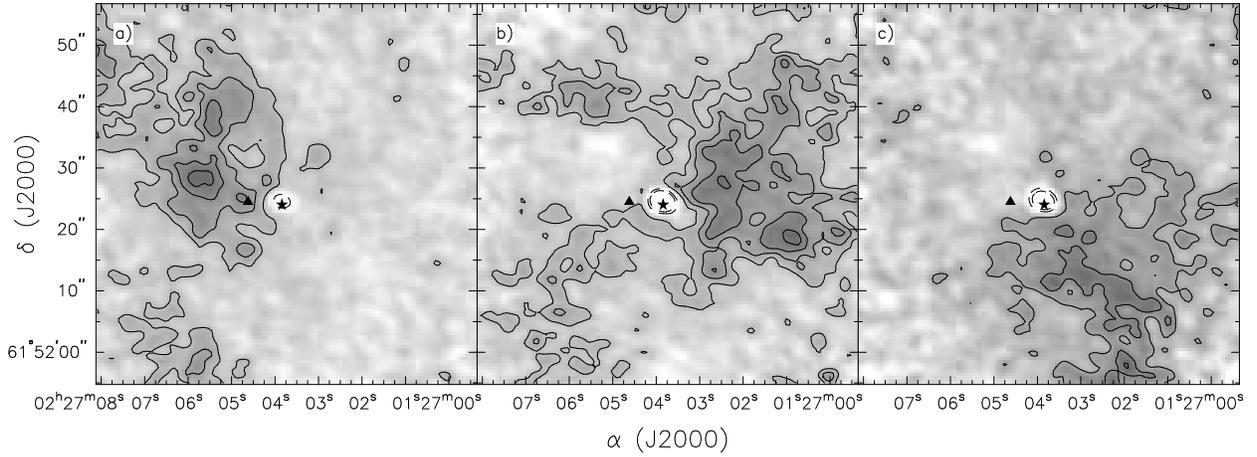}
\end{center}
\caption{Integrated line maps of CN velocity components in
  W3(OH). These are the three distinct velocity components present in
  the CN window. Each map is averaged over the two hyperfine
  components present. The figure \emph{(a)} is averaged from -52.7 to
  -48.5 km s$^{-1}$, figure \emph{(b)} from -48.2 to -46.3 km
  s$^{-1}$, and figure \emph{(c)} from -46.3 to -43.1 km
  s$^{-1}$. Notable features include the dense clumps in the
  north-east corner of \emph{(a)}, and the continuum source which can
  be seen in absorption. The contours are 2, 4, 6, and 8 times the noise
  level of about 0.03 Jy beam$^{-1}$. The $\bigstar$ and
  $\blacktriangle$ represent the peak continuum positions of W3(OH)
  and the Turner-Welch object, respectively.
\label{cnchan}}
\end{figure}

Compared to the other molecules, the CN emission appears to be diffuse
but ``clumpy.'' Under closer scrutiny, it is much more complex. Our 8
MHz spectral window is centered at 113.490982 GHz, the frequency of
CN's strongest hyperfine component. This frequency and window size was
chosen to allow us to simultaneously image a second hyperfine
component of the CN N=1-0 transition at 113.488 GHz in the same
spectral window. These two hyperfine components are separated by 7.9
km s$^{-1}$. Since this window covers approximately 21 km s$^{-1}$ of
velocity space, we expect to fully resolve and image both
lines. Multiple velocity components in the image make it difficult to
determine which hyperfine component some emission belongs to. However,
we have been able to identify at least three distinct velocity
components, representing separate regions around W3(OH) (Figure
\ref{cnchan}).

The north-west and south-west components appear to be very
diffuse. They are centered at approximately -47.25 km s$^{-1}$ and
-44.5 km s$^{-1}$, respectively. It is difficult to completely isolate
these two velocity components due to blending and overlap of the
hyperfine lines. The maps in Figure \ref{cnchan} reflect some of this
``blending.'' It occurs because the second hyperfine component of one
velocity partially overlaps the primary hyperfine component of the
other velocity. For example, the second and third maps in Figure
\ref{cnchan} are cross-contaminated. This is easily visible when the
positions of the emission in the two maps are compared; they have a
significant spatial overlap region.

\begin{figure}
\begin{center}
\includegraphics[scale=.75, angle=-90]{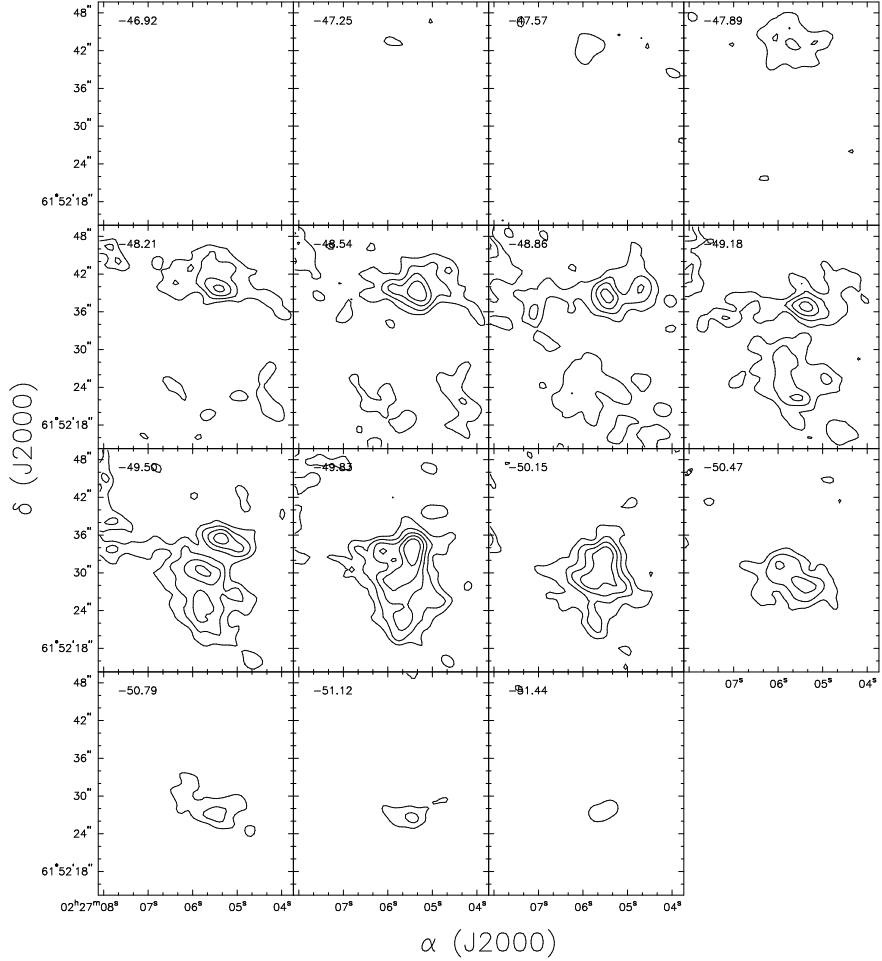}
\end{center}
\caption{Channel maps of the CN clump in the north-eastern corner of
  the CN map. Displayed are 15 channels from -46.92 km s$^{-1}$ to
  -51.44 km s$^{-1}$. There are two visible ``lobes'', the northern
  one which peaks at -48.64 km s$^{-1}$, and the southern one which
  peaks at -49.83 km s$^{-1}$. The countour levels are 5, 8, 11, 14, 
  17 times 0.05 Jy beam$^{-1}$.
  \label{outflowcm}}
\end{figure}

\begin{figure}
\begin{center}
\includegraphics[scale=0.75, angle=-90]{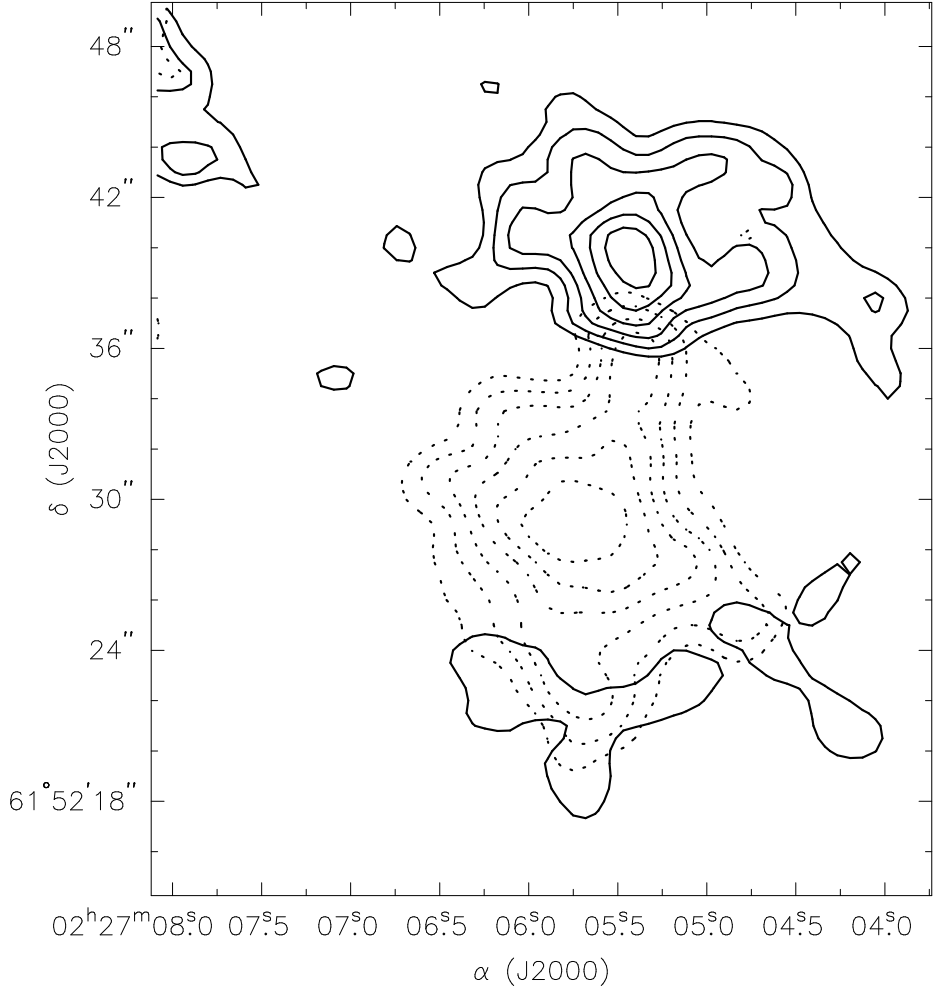}
\end{center}
\caption{Averaged map of the two ``lobes'' of the outflow-like
  feature. The emission represented by solid contours is averaged from
  6 channels between -47.6 km s$^{-1}$ and -49.2 km s$^{-1}$. The
  emission represented by dotted contours is averaged from 6 channels
  between -49.2 km s$^{-1}$ and -50.8 km s$^{-1}$. While this does
  look similar to an outflow, it could also be bulk rotation of a
  clump of gas. The contour levels of both objects are 4, 5, 6, 7, 8,
  9 times 0.06 Jy beam$^{-1}$.
  \label{outflowcombined}}
\end{figure}

The north-eastern component is more complex. In addition to containing
diffuse emission, multiple denser clumps are visible. The significance
of some of these clumps are not clear from the integrated line
map. Between the larger southernmost clump and the smaller, northern
clump are a string of small, presumably unresolved clumps only seen in
a single channel each. When viewing an averaged line map, many of
these features are spatially smoothed and are not prominent (as in
Figure \ref{cnchan}). The densest portions of this region are
concentrated in two lobes, not unlike the typical signature of an
outflow. Figure \ref{outflowcm} shows channel maps of the 15 channels
around the strongest hyperfine component. There does not appear to be
any blending as was seen in the south west component, so we are
confident that this channel map is not contaminated. An averaged
contour plot (Figure \ref{outflowcombined}) shows the physical
comparison between the two ``lobes'' of the emission. The top (red
shifted) lobe is averaged from the six channels from -47.6 km s$^{-1}$
to -49.2 km s$^{-1}$, while the bottom (blue shifted) lobe is averaged
from the six channels from -49.2 km s$^{-1}$ to -50.8 km
s$^{-1}$. While this does have the signature of a typical outflow, it
is not accompanied by any coincident emission from any of the other
species, nor is there any accompanying IR emission source visible in
publicly available IR catalogs. Due to the extent of the low level
emission around the whole region, and lack of evidence of an outflow
generating source, it is more likely that this is not an outflow, but
rather bulk rotation of a clump of gas.

\subsection{HCN, HCO$^+$, C$^{18}$O}

\begin{figure}
\begin{center}
\includegraphics[height=\linewidth, angle=-90]{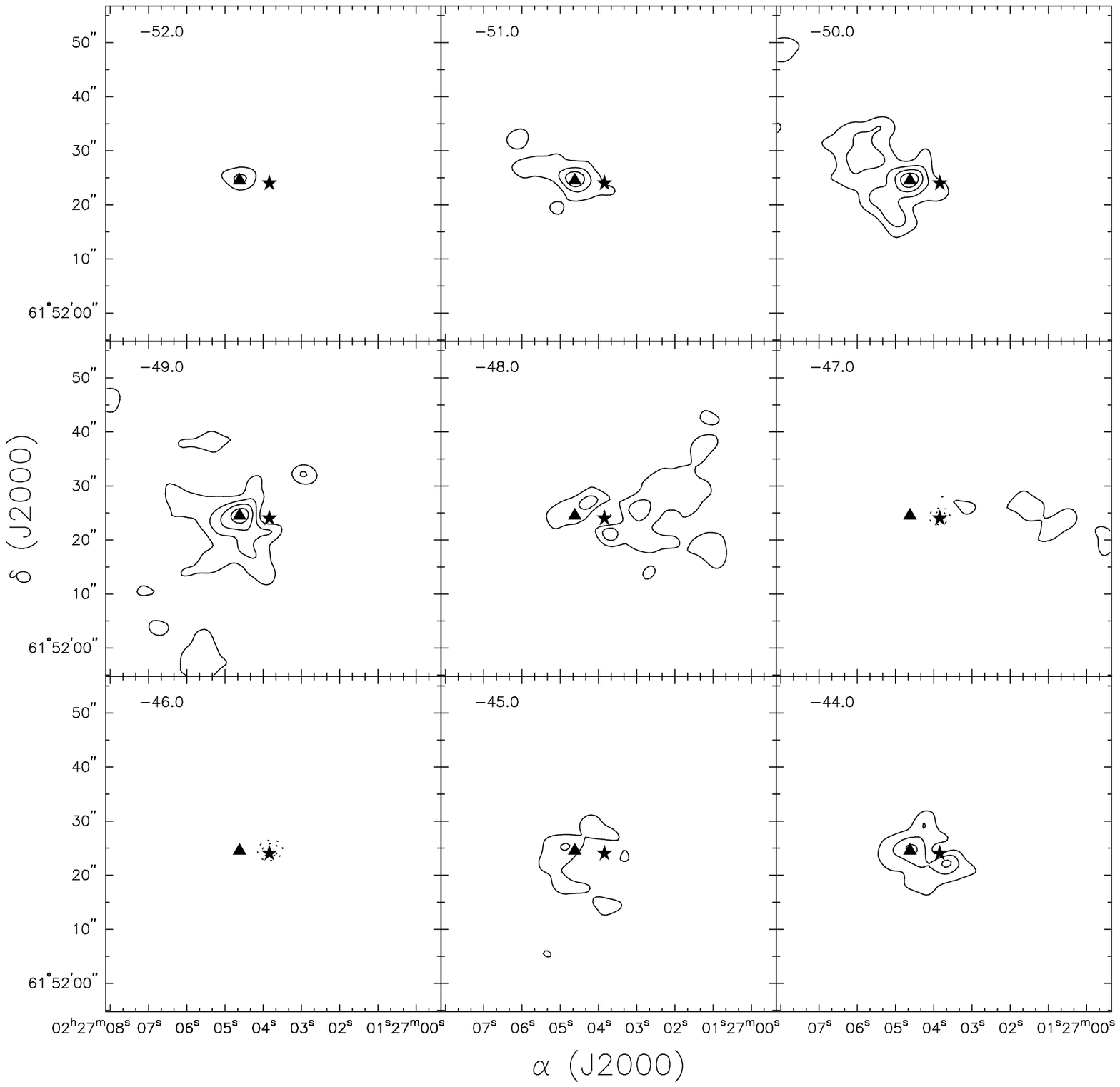}
\end{center}
\caption{HCN Channel maps, rebinned to 1 km s$^{-1}$ velocity
  increments. The velocity scale is centered on the strongest HCN
  hyperfine component at 88.631 GHz. The spacial resolution of this
  map is 3.4 x 2.6 arcsec. The contour levels are -80, -40, 20, 40,
  60, 80, 100 times 0.0287 Jy beam$^{-1}$. The $\bigstar$ and
  $\blacktriangle$ represent the peak continuum positions of W3(OH)
  and the Turner-Welch object, respectively.
\label{hcnchannel}}
\end{figure}

\begin{figure}
\begin{center}
\includegraphics[height=\linewidth, angle=-90]{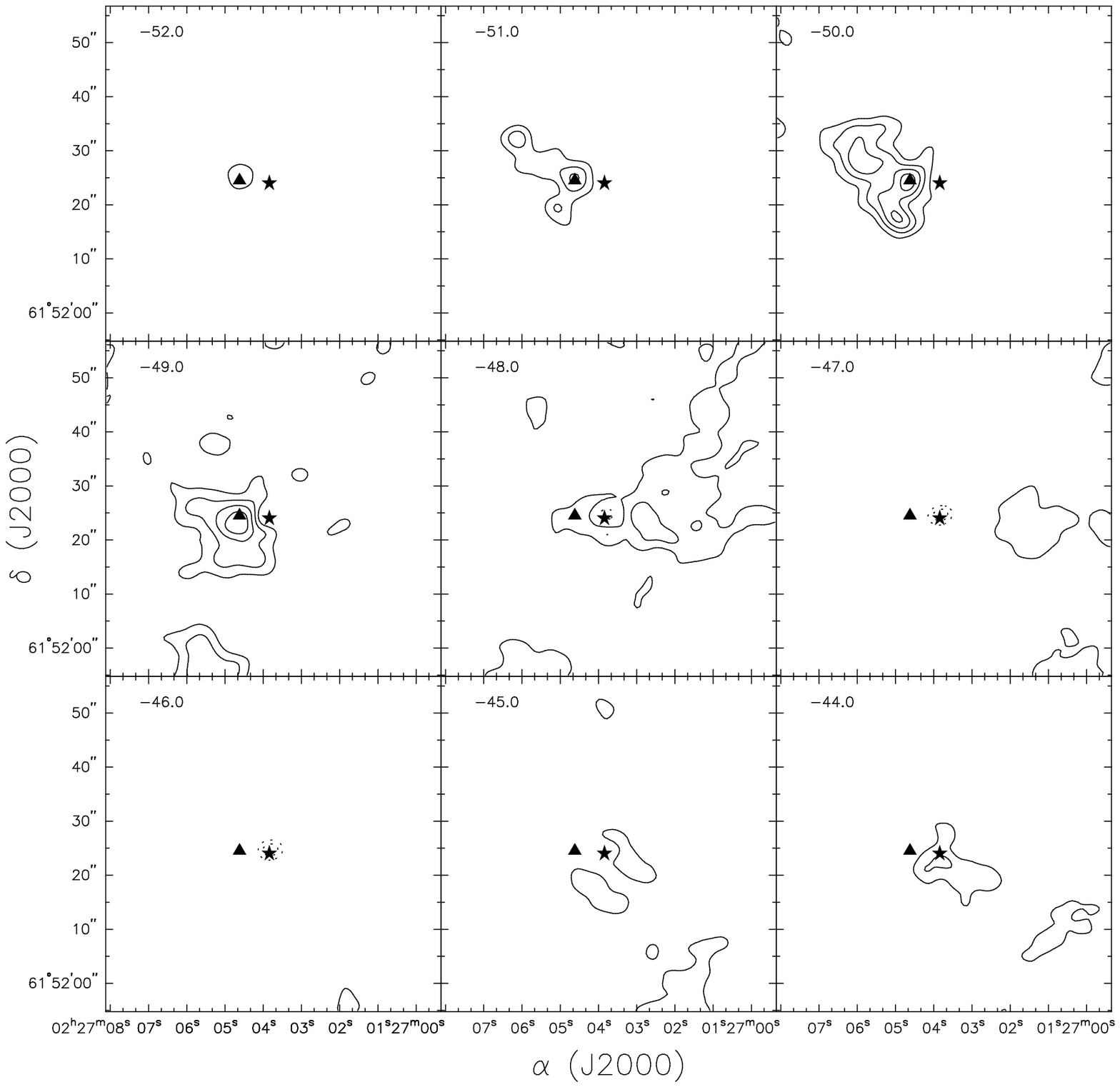}
\end{center}
\caption{HCO$^+$ Channel maps, rebinned to 1 km s$^{-1}$ velocity
  increments. The spacial resolution of this map is 3.5 x 2.7
  arcsec. The contour levels are -80, -40, 20, 40, 60, 80, 100 times
  0.0296 Jy beam$^{-1}$. The $\bigstar$ and $\blacktriangle$ represent
  the peak continuum positions of W3(OH) and the Turner-Welch object,
  respectively.
\label{hco+channel}}
\end{figure}

\begin{figure}
\begin{center}
\includegraphics[height=\linewidth, angle=-90]{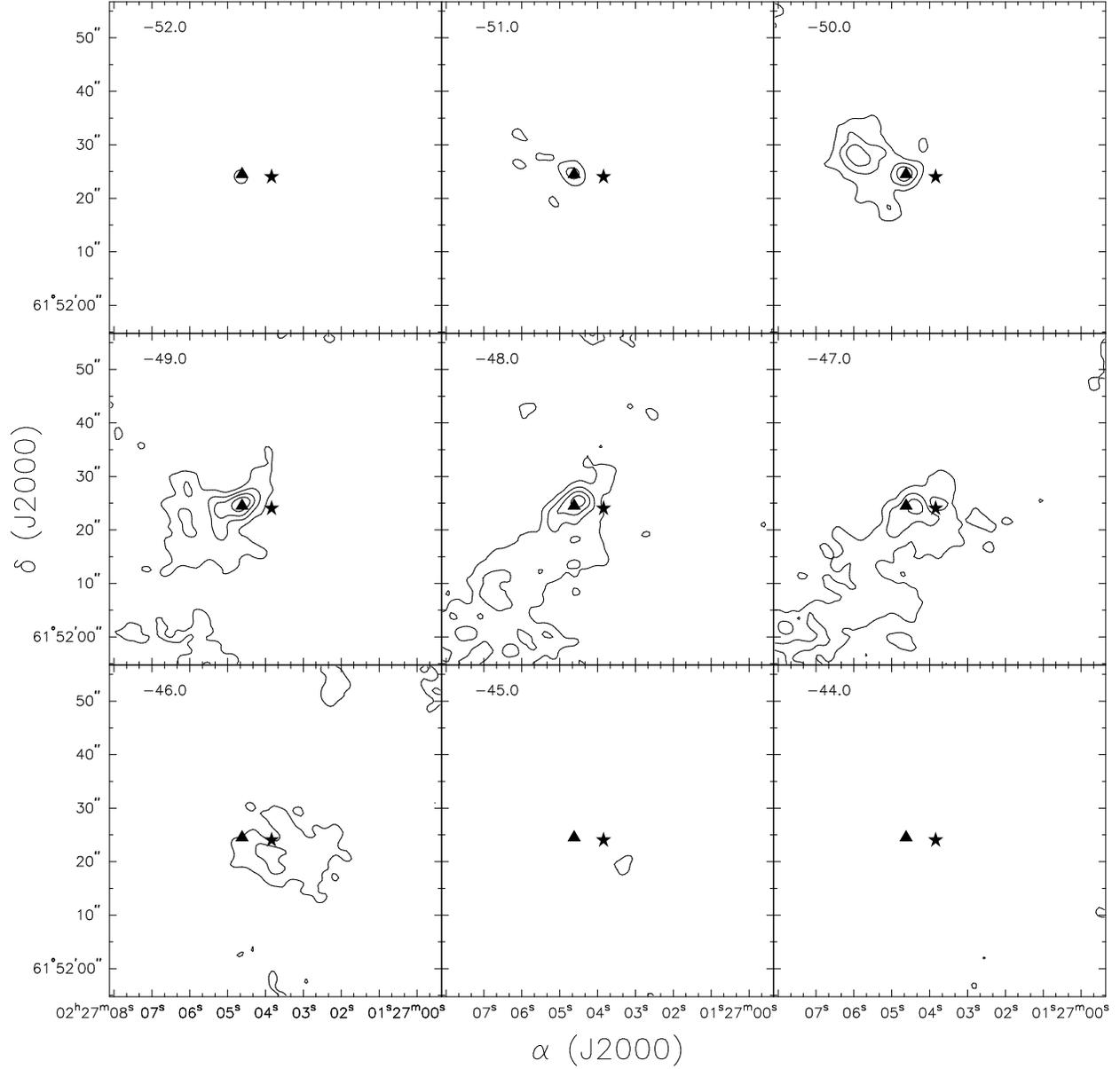}
\end{center}
\caption{C$^{18}$O Channel maps, rebinned to 1 km s$^{-1}$ velocity
  increments. The spacial resolution of this map is 3.1 x 2.5
  arcsec. The contour levels are 20, 40, 60, 80, 100 times 0.00988 Jy
  beam$^{-1}$. The $\bigstar$ and $\blacktriangle$ represent the peak
  continuum positions of W3(OH) and the Turner-Welch object,
  respectively.
\label{c18ochannel}}
\end{figure}

\begin{figure}
\begin{center}
\includegraphics[height=\linewidth, angle=-90]{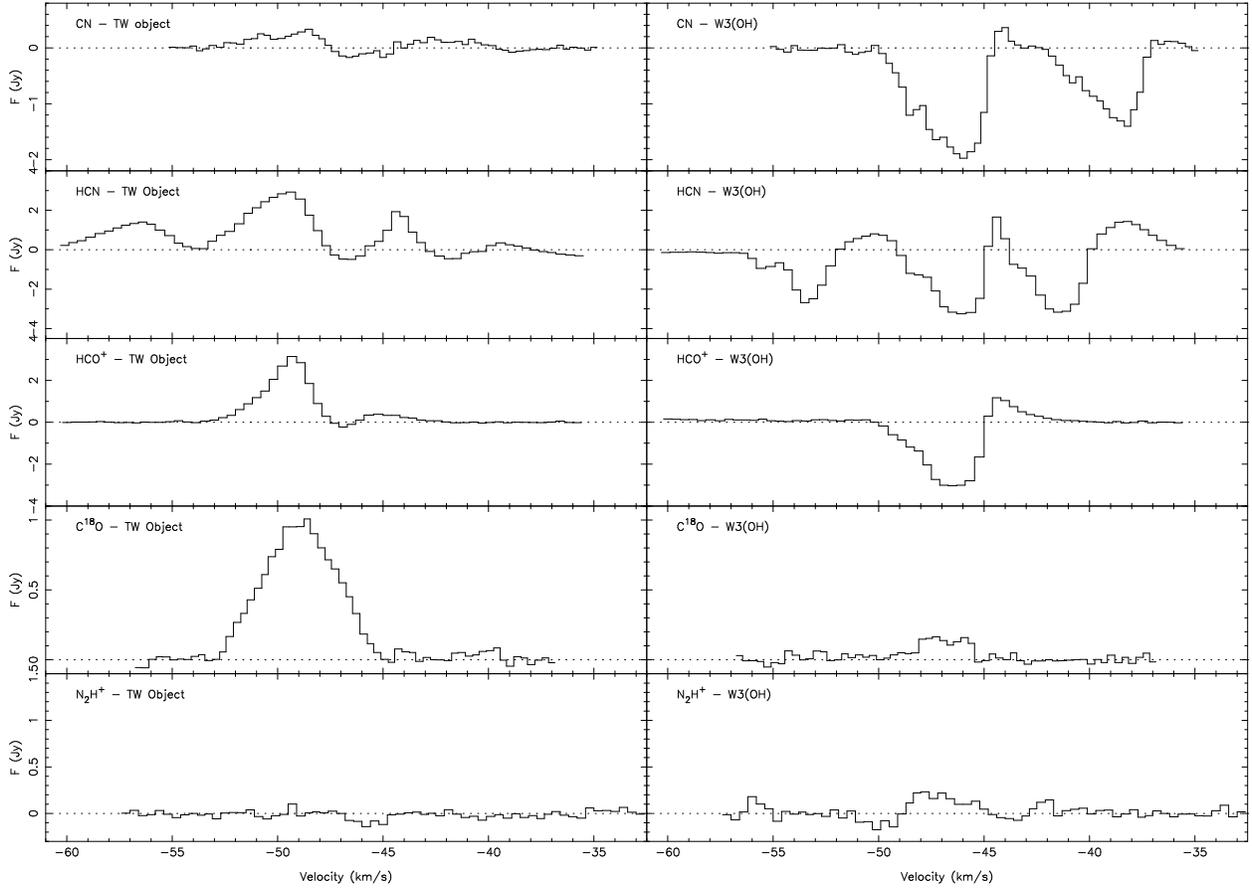}
\end{center}
\caption{Sample spectra for each species, towards both the continuum
  source (W3OH) and the Turner-Welch object. For the CN, HCN, and N$_2$H$^+$
  spectra, multiple hyperfine components are present. The velocity
  scale for all of these corresponds to that of the strongest
  hyperfine component. Three of the sources can be seen in absorption
  towards the continuum source, while the other two have little to no
  emission.
\label{spectra}}
\end{figure}

The maps for HCN, HCO$^+$, C$^{18}$O all appear to be tracing the same
material, which does not coincide with either the CN or N$_2$H$^+$
emission (see Figures \ref{hcnchannel}-\ref{c18ochannel} for channel
maps).  However, the peak position of these spectral lines coincides
with the HCN detection of the Turner-Welch object first reported in
\citet{turner84}. At 2.5\arcsec\ resolution we do not appear to be
resolving the object; previous high resolution (sub-arcsecond) studies
have been conducted in order to resolve this object in the continuum
\citep{chen06}, which resulted in the detection of multiple
cores. Previous HCO$^+$ studies \citep{wink94} had similar spatial
resolution to our data, however, our data has significantly better
signal-to-noise and samples material that was resolved out in the
previous study. In the HCO$^+$ spectrum of the Turner-Welch object, a
weak, secondary velocity component at -44 km s$^{-1}$ is visible
(Fig. \ref{spectra}). This velocity component is not reported in any
previous studies. The HCN spectra towards the Turner-Welch object
shows an unexpected peak at an apparent velocity of -39 km s$^{-1}$
(Fig. \ref{spectra}); comparing relative positions of these two peaks,
the anomalous HCN peak would be the F=0-1 transition of a secondary
velocity component at -44 km s$^{-1}$. The other hyperfine lines from
this velocity component are being masked by the hyperfine lines from
the primary velocity component. This can only happen if the separation
between the hyperfine lines is approximately the separation between
the two velocity components. The C$^{18}$O line may also show this
secondary component, however, it cannot be confirmed due to the
relatively low SNR of the C$^{18}$O emission.

The C$^{18}$O map does not show any absorption towards the continuum
source, unlike HCN, HCO$^+$, and CN. This is consistent with
observations done by \citet{wink94}, in which they surmised that
the line excitation temperature of HCO$^+$ is significantly lower than
the kinetic temperature of 90 K; but the C$^{18}$O excitation
temperature is not low enough to produce absorption. However, our
observed absorption in other lines is significant and can yield
information about the regions around the continuum source. The HCO$^+$
spectra, seen in absorption (Figure \ref{spectra}), looks similar to
an inverse P-Cygni profile which could indicate an expanding
shell-like structure or a strong stellar wind.  The multiple hyperfine
components of HCN make it difficult to tell which gas is causing the
absorption; however, it has the same central velocity and line width
as the HCO$^+$ absorption indicating that both features are being
generated from the same region. The CN absorption also appears to have
similar spectral features as HCN; however the red-shifted CN emission
feature is at a level of less than 1$\sigma$ of the noise level, and
cannot be considered as a positive detection.

Another notable feature of the continuum absorption lines is their
shape. There is a double trough shape seen in HCO$^+$, the strongest
hyperfine component of HCN, and in CN (these spectra are centered on
the strongest hyperfine component). There is also some evidence of
this feature in the other hyperfine components; however they are
weaker and, as a result, noisier. According to the results of
\citet{wink94}, within the UCHII region of W3(OH) there is an embedded
O7 star. It is very likely that W3(OH) is a region surrounding a
bright massive star embedded in a dense cloud with a strong solar wind
clearing out the region around the star. In looking at a velocity
moment map in HCO$^+$ (HCO$^+$ was chosen due to its lack of hyperfine
components and contamination), there appears to be an East-West
velocity gradient on the order of 3 km s$^{-1}$. Since W3(OH) is
unresolved in these maps, higher resolution data of W3(OH) are
required to quantify and describe this effect (see below).

\subsection {N$_2$H$^+$}

Figure \ref{n2h+channel} shows N$_2$H$^+$ channel maps. If N$_2$H$^+$
were tracing the same dense material as CN, we would expect to see
very similar emission (and absorption) spectra. Most notable is the
lack of an absorption feature towards the continuum source, or
alternatively, of any emission in N$_2$H$^+$ at the continuum
position. Also, the rotation feature in CN noted above is not seen in
N$_2$H$^+$. The N$_2$H$^+$ emission appears to be concentrated in the
south western region of the map, roughly coincident with one of the
velocity components of CN; however, this is where the similarity
ends. In the N$_2$H$^+$ spectra (Figure \ref{spectra}), only 3 peaks
appear when there are 7 hyperfine lines within our window. The central
peak, at about -47.5 km s$^{-1}$, contains 3 hyperfine components
which, having only 0.487 MHz (1.56 km s$^{-1}$) between them, are
blended together. The peak centered at -42.5 km s$^{-1}$ contains
another three hyperfine components whose transition frequencies all
lie within 0.432 MHz (1.39 km s$^{-1}$) and would also appear
blended. The last hyperfine component is seen at -56 km s$^{-1}$ and
is the only one that is not blended. Our N$_2$H$^+$ map contains some
similarities to the ammonia (NH$_3$) map presented by \citet{wilson93}
and \citet{tieftrunk98}. The major difference between the two is the
lack of N$_2$H$^+$ emission around the Turner-Welch object. The major
similarity comes from an east-west elongated clump present in the
northeast corner of both the N$_2$H$^+$ emission and the NH$_3$
emission. This stands out since none of the other molecules show any
sign of emission from this region. The closest emission is the north
lobe of the possible CN outflow. It is unclear if these regions are
related.

\begin{figure}
\begin{center}
\includegraphics[height=\linewidth, angle=-90]{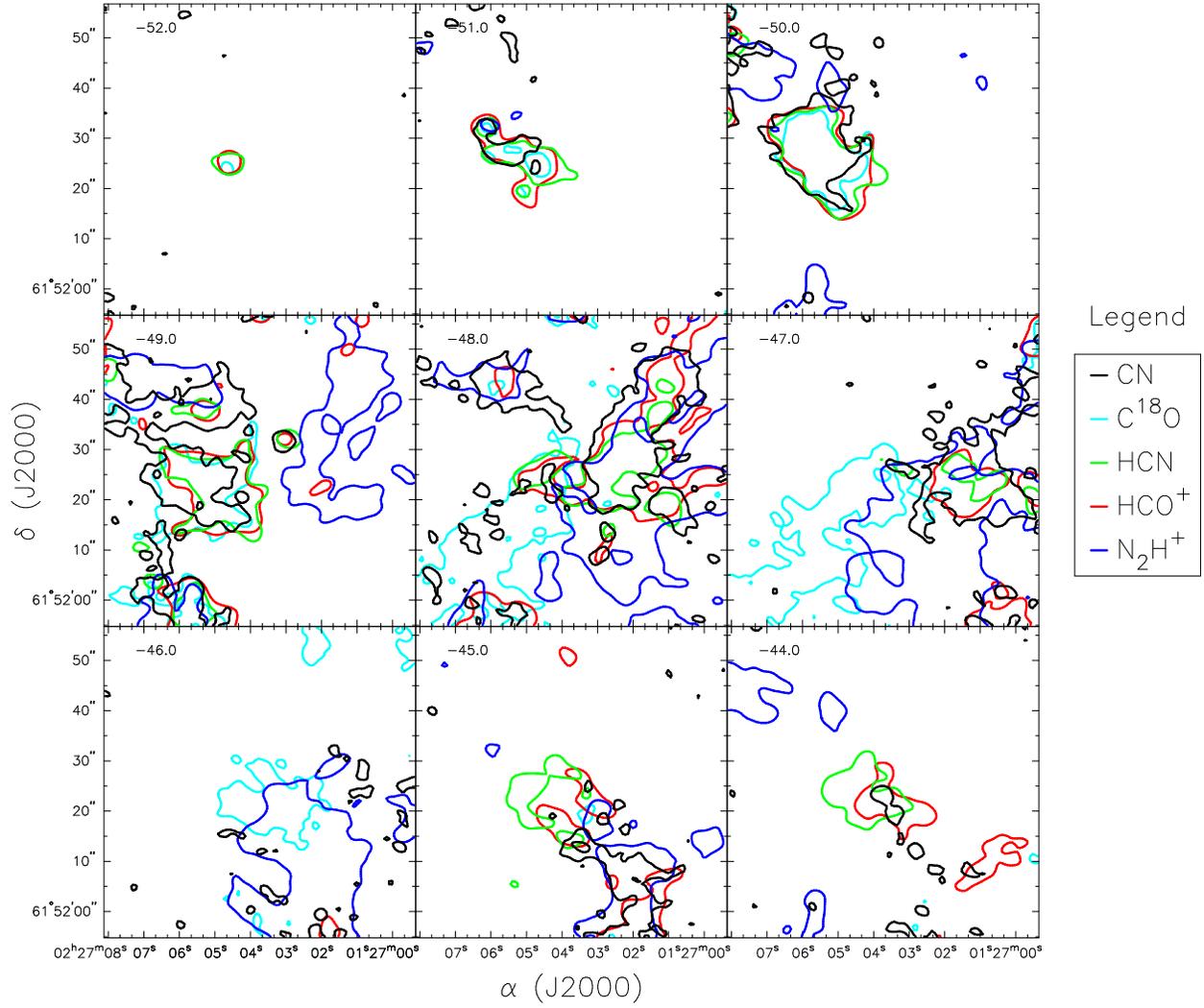}
\end{center}
\caption{Channel map showing the 20\% contour level of all the mapped
  species. \label{combined}}
\end{figure}

It is possible that the regions of CN emission much more closely
matches the regions of N$_2$H$^+$ than is readily apparent, even
though the spatial peaks do not coincide. Since CN is significantly
weaker than N$_2$H$^+$, it is possible that emission, particularly in
the lower west side of the map that is detectable in N$_2$H$^+$, is
significantly below the noise threshold in CN. To emphasize this fact,
which may not be noticeable in the individual channel maps, we
constructed a channel map with 20\% contours from all five detected
species (Fig. \ref{combined}). Most noticeable in the -48 km s$^{-1}$
and -47 km s$^{-1}$ panels, CN (with several other of the species)
does seem to trace much of the N$_2$H$^+$ emission. In the -51 km
s$^{-1}$ and -50 km s$^{-1}$ panels, it appears that the N$_2$H$^+$
emission borders on the regions with emission in the other species. If
we assume that the two molecules do trace the same density gas, it is
quite possible that some of these regions are undergoing a chemical
reaction that selectively annihilates N$_2$H$^+$.

\subsection{B-array Observations}

From the observations described above, several significant features
warranted even higher resolution observations.  The continuum source
is seen in absorption in some species; however, it remains
unresolved. CARMA's B-array should be able to probe the region at
significantly higher resolution than in our previous data. This will
additionally allow us to study the velocity structure of gas near the
continuum source. A velocity map of HCO$^+$ shows an East-West
velocity gradient of $\sim$ 3 km s$^{-1}$. HCO$^+$ was chosen
specifically since it does not have multiple hyperfine components
which, due to blending, could pollute the velocity map, and shows
absorption towards the continuum source. In addition, with these data
we can see if we can resolve the individual cores of the Turner-Welch
object. Previous high resolution continuum maps have shown evidence of
multiple cores within the Turner-Welch object which we would expect to
also see in corresponding high resolution spectral line maps. We
previously choose to map the region with the C, D, and E arrays to
study large scale features; with B-array alone we are able to map
features that require higher spatial resolution. If we combined the
B-array data with the rest, we will produce a map with a synthesized
beam that is larger than the beam of the B-array data alone, and
larger than some of the smallest features, obscuring them. In
addition, the necessity to self-calibrate the data and issues with
gain calibration (discussed in detail below) make a separate analysis
more informative.

Self calibration results in several issues that need to be
overcome. W3(OH)'s physical size is 0.01 pc in diameter
\citep{kawamura98} which corresponds to an angular size of $\sim$
1\arcsec. This means that many of the longer baselines in B-array will
be resolving the W3(OH) continuum source. Data from these baselines
will make it difficult to self calibrate the data as it assumes a
point source model for the source (or requires an accurate model of
the source features which we do not have). Running self-cal on the
full dataset (including resolved baselines) generates a solution that
is appropriate for the data up until a UV-radius of 130k$\lambda$, at
which the solution fails. This UV-radius corresponds to a physical
size of:
\begin{eqnarray*}
\Theta & \sim & \frac{\lambda}{D} \\
       & \sim & \frac{\lambda}{2 * \mathrm{UVr}} \\
       & \sim & 0.79\arcsec
\end{eqnarray*}
Therefore running self-cal on the full dataset assuming a point source
will result in large errors in the selfcal solution due to long,
resolved baselines. If we ``cut'' the dataset so we only use the data
with UV-radius $<$ 130k$\lambda$ for use in calculating the phase
solution, we will not have this issue.

The selfcal technique fits the phase center of the data to the
brightest point source in the map. Our C, D, and E array data is
slightly offset in position from the pointing center, so this will
result in a slight position offset between the two maps. The can be
corrected for by manually fitting and entering the center position of
the continuum source from our C, D, E maps, however this is not
necessary since we will not be directly combining our B-array data
with the rest, since the resulting maps at full angular resolution
would have a very low signal-to-noise ratio.

This self-calibration technique also poses several problems with
regards to flux and amplitude calibration. Several dishes primarily
have long baselines, with only a few baselines that fall under our
UV-radius limit of 130k$\lambda$. While this worked well for phase
calibration, the results of amplitude selfcal was very poor. Many of
the amplitude gains, which should be flat and close to unity, were
noisy and had many datapoints corresponding to unphysical gains that
directly translates to poor image quality. The few remaining baselines
also tend to have low signal to noise which further increases the
problem.

In the attempt to apply correct gain calibration, several other
techniques were tested to amplitude calibrate the data. We attempted
to amplitude calibrate our data using the dedicated phase calibrator
measurements and transferring the resulting solution to our
source. This also failed, and while it produced better results than the
amplitude selfcal technique described above, the high resolution, low
intensity features were washed out in the final maps. This is most
likely due to the gain calibrator being 10$^o$ from our source, far
enough that the telescopes are viewing different enough atmosphere to
negatively affect the calculated gains. This problem does not affect
our C, D, E array data because the maximum baseline length is
significantly smaller in these arrays and are not as sensitive to
atmospheric fluctuations as the longer B-array baselines. Since these
several technique turned out poor results, it was decided not to use
amplitude (flux) calibration on the B-array data, which is not an
uncommon result when using selfcal. This is another reason why we do
not combine this data with our C, D, E data as we would need an
accurate gain solution to provide proper weights to combine the
datasets.

\subsection{Continuum}

\begin{figure}
\begin{center}
\includegraphics[height=\linewidth, angle=-90]{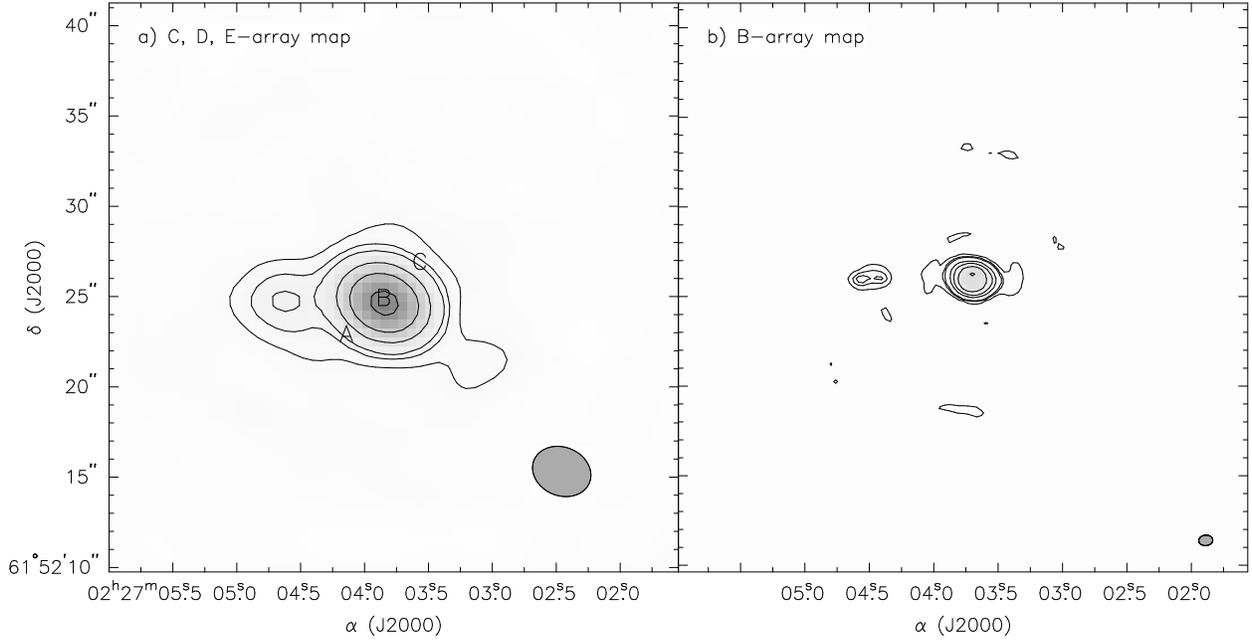}
\end{center}
\caption{Continuum maps centered at 112 GHz. Panel a) has contour
  levels of 8, 16, 25, 64, 128, and 256 times 0.009 Jy beam$^{-1}$.
  Panel b) has contour levels of 3, 7, 10, 32, 64, 128, and 256 times
  0.003 Jy beam$^{-1}$.  The range in contour levels are to show the
  dynamic range of features visible in these maps. a) has a beam size
  of 3.3\arcsec x 2.7\arcsec, b) has a beam size of 0.77\arcsec x
  0.59\arcsec. Positions A-C marked in the first panel represent the 
  positions of spectra in Figure \ref{hco+cuts}.
\label{continuumall}}
\end{figure}

Figure \ref{continuumall} represents the combined C, D, E array and
B-array only continuum maps. Each map was produced using
multi-frequency synthesis (MFS) on two 500 MHz windows, one in the
N$_2$H$^+$ band (average frequency of 91.2 GHz, not shown), and one in
the CN band (average frequency of 112 GHz, shown). In Figure
\ref{continuumall}a, neither the Turner-Welch object nor W3OH are
resolved. The peak flux of W3OH decreases by a factor of 2 from 91.2
GHz to 112 GHz, while the center of the Turner-Welch object increases
in flux by a factor of two over the same range. In the B-array only
data (Figure \ref{continuumall}b), we resolve two individual clumps
within the Turner-Welch object and slightly resolve structure within
W3OH (note the off-center position of the highest level contour in
Fig. \ref{continuumall}b). This structure supports the findings of
\citet{chen06}. We additionally see the same change in relative fluxes
in the B-array maps as we saw in the C, D, E array maps, over the same
frequency range.

\section{Discussion}

\subsection{W3(OH) Absorption Feature}

\begin{figure}
\begin{center}
\includegraphics[height=\linewidth, angle=-90]{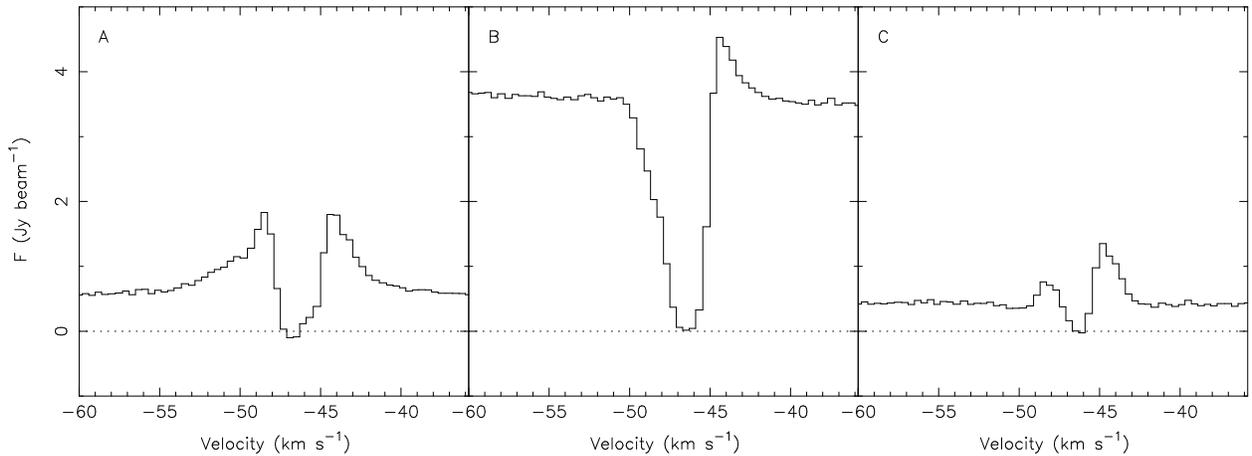}
\end{center}
\caption{Panels A-C are HCO$^+$ spectra of W3(OH) at three positions
  with no continuum subtraction. This shows the change in emission and
  absorption spectra in HCO$^+$ across the W3(OH) Ultra-Compact HII
  region. Panel A is 2\arcsec\ SE of the W3(OH) center position, and
  Panel C is 2\arcsec\ NW of the W3(OH) center position. The positions
  of spectra A-C are marked on Figure \ref{continuumall}a for
  reference.
\label{hco+cuts}}
\end{figure}

As stated above, W3(OH) itself is a UCHII region with an embedded O7
star. Our data show that several of the molecular tracers we have
observed can be seen in absorption towards this source. The nature and
structure of this colder absorbing gas may be able to give us some
insight into the properties of W3(OH). Figure \ref{hco+cuts} shows
three spectra sampled across the continuum source in HCO$^+$. HCO$^+$
was chosen for this since it does not have any hyperfine components
that could cause line profile confusion.  C$^{18}$O, while also not
having any hyperfine components, was not chosen to trace this feature
as it is not seen in emission or absorption towards W3(OH). The
composite emission and absorption spectra are consistent with a dense,
hot region (containing HCO$^+$) surrounded by one or more cold,
less-dense layers which also contain HCO$^+$. The absorption feature
is evidence of cold, optically thick gas which is at a velocity of -46
km s$^{-1}$ and has a FWHM of 4 km s$^{-1}$. In looking at
non-continuum subtracted spectra (Figure \ref{hco+cuts}b), this
absorption line is saturated and extends completely down to zero
flux. This implies that this cold region is in front of the emission
region and the continuum source. The absorption feature is also seen
in CN and HCN, however, it is seen slightly in emission in
C$^{18}$O. The C$^{18}$O emission feature is also centered at -46 km
s$^{-1}$ and has a FWHM of 4 km s$^{-1}$, the same as the HCO$^+$
absorption. It is very likely that it is from the same region,
implying that the gas density is sufficient to raise the excitation
temperature of the C$^{18}$O line above that of the brightness
temperature of the continuum source so the line appears in emission,
while the higher critical densities of the HCO$^+$, HCN, and CN
transitions lead their excitation temperatures to be less than that
brightness temperature, so they are seen in absorption. To note:
Figure \ref{hco+cuts} is resolution limited since it is formed with C,
D, and E array data and its three positions only cover an area of
1.5-2 beamwidths. This was part of the motivation to acquire B-array
data as the same region would cover 5-8 beamwidths in a B-array map.

\begin{figure}
\begin{center}
\includegraphics[height=\linewidth, angle=-90]{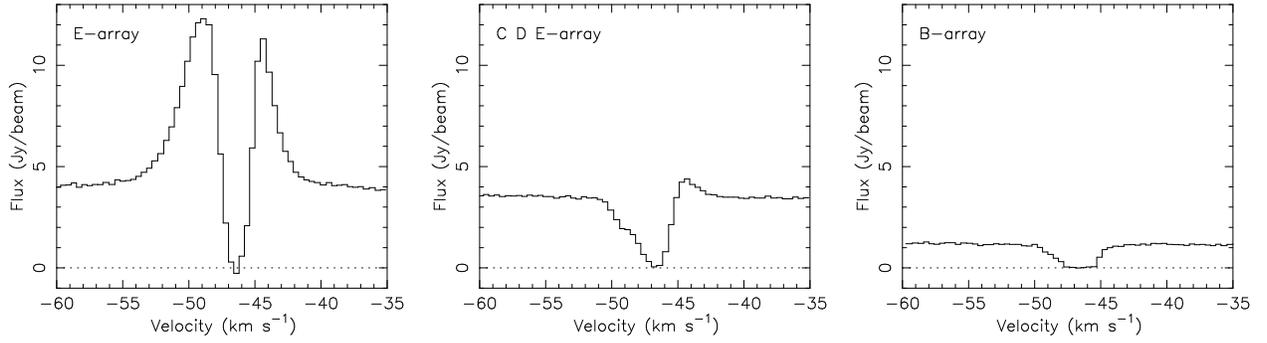}
\end{center}
\caption{Comparison between spectra towards the center of the
  continuum source in a) E-array only, b) combined C, D, E array maps,
  and c) B-array only. In all three instances, the spectra are seen in
  absorption at the same velocities and with roughly the same
  linewidth which supports a model of a cold dark, optically thick
  cloud in front of the continuum source and emission regions in our
  map.
\label{hcospectra}}
\end{figure}

From the B-array spectra (Figure \ref{hcospectra}c), we only see
HCO$^+$ in absorption against the continuum emission. The lack of
emission either means that it is being resolved out at this high
spatial resolution, or that the beam is completely filled by continuum
emission, so we see no emission from the sides of the beam. In either
case, this means that the emission does not come from the same gas
that produces the absorption. To support this, we produced spectra
that only includes E-array (largest beam size, shortest baselines) in
order to inspect large scale emission. The E-array data (Figure
\ref{hcospectra}a) shows emission towards the continuum between -43 km
s$^{-1}$ and -50 km s$^{-1}$ with a self absorption feature at -46 km
s$^{-1}$. This high velocity gas is not seen in C$^{18}$O and
therefore seems to be hot, optically thin gas that is behind the
continuum source.

Therefore, we propose a multi-layered model of the region around
W3(OH). Behind the UCHII region is a high velocity, hot, large spatial
scale, optically thin region which appears to peak in intensity
towards the Turner-Welch object (and may very well be associated with
it). The UCHII region is optically thin, and has a small spatial scale
of about 0.7\arcsec in diameter which is extremely bright in the
continuum, enough such that the narrow band 2 MHz windows are
contaminated by continuum emission. In front of the continuum is a
very cold, optically thick region at a velocity of -46 km s$^{-1}$
which we primarily see in absorption, but we see slightly in emission
in C$^{18}$O.

\subsection{CARMA Flux}

\begin{figure}
\begin{center}
\includegraphics[height=\linewidth, angle=-90]{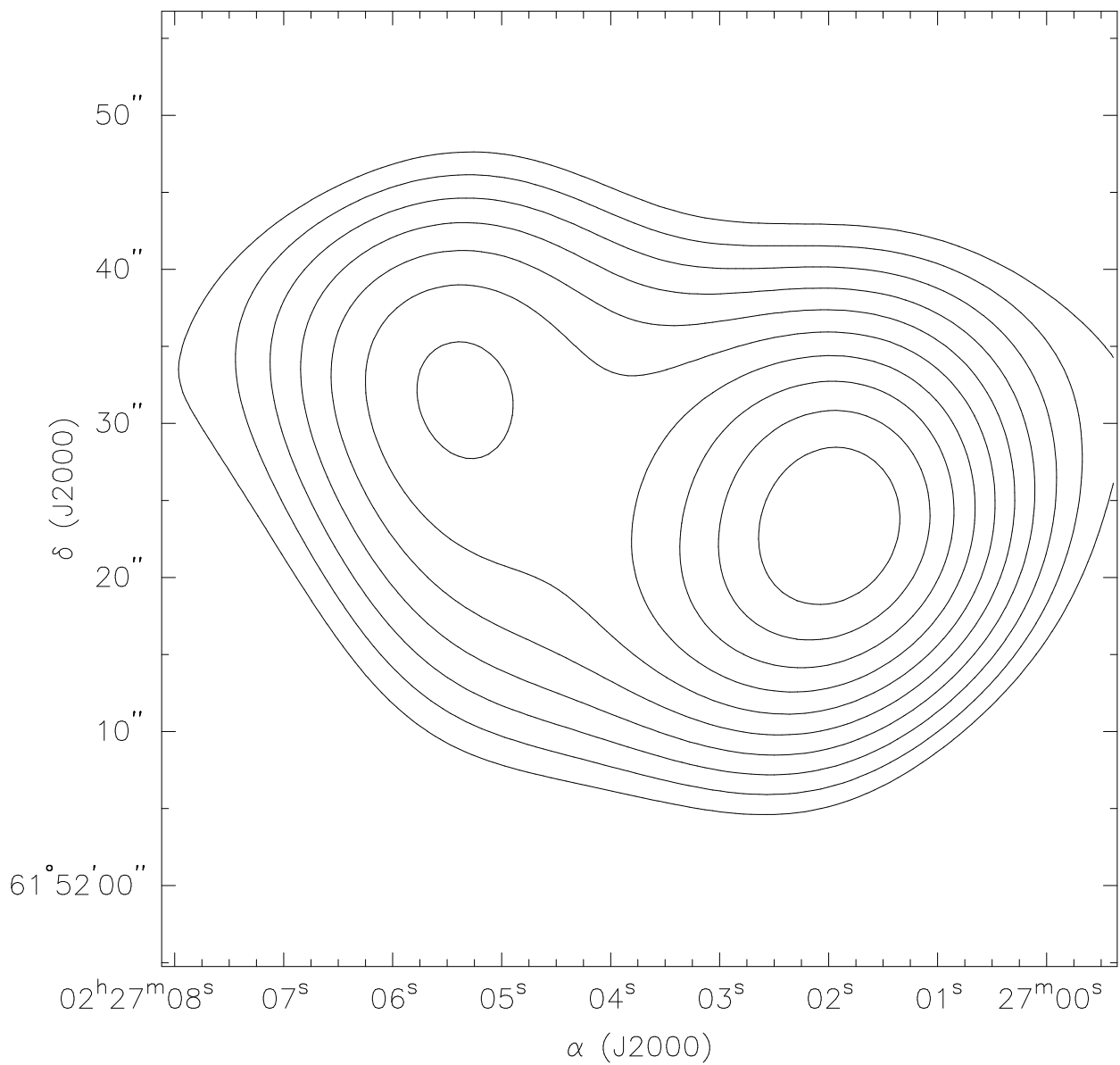}
\end{center}
\caption{Smoothed CARMA CN map to the same resolution as the IRAM
  measurements (23$\arcsec$). This was produced by convolving a
  23\arcsec\ Gaussian beam with the integrated line map seen in
  Fig. \ref{w3ohmap}. The contour levels are in increments of 0.158 
  Jy beam$^{-1}$ from 1.89 Jy beam$^{-1}$ to 3.47 Jybeam$^{-1}$.
\label{iramcomp}}
\end{figure}

\begin{figure}
\begin{center}
\includegraphics[height=\linewidth, angle=-90]{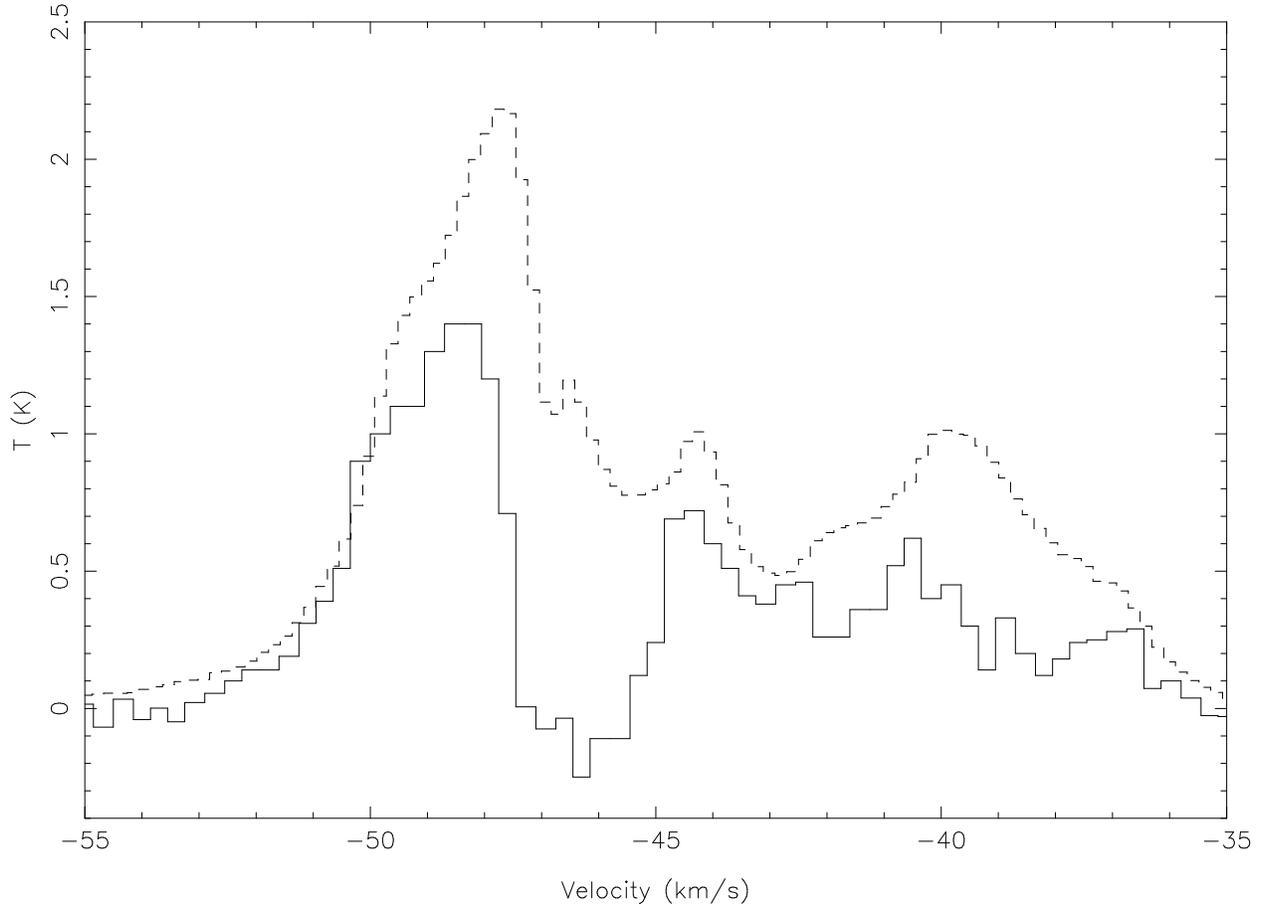}
\end{center}
\caption{CARMA spectrum overlaid with the IRAM spectrum at a point
  near the IRAM pointing center. Note that there are two CN hyperfine
  lines in these spectra, separated by about 8 km s$^{-1}$. The
  velocity scale is for the stronger hyperfine line (at $\sim$48 km
  s$^{-1}$). The intensity scale is estimated by comparing the main
  beam efficiencies of IRAM and CARMA, however, there is an inherent
  20\% error in the CARMA flux measurements. From this, we estimated
  that CARMA resolves $\sim$65\% of the CN emission within the IRAM
  beam.
\label{iramcom2}}
\end{figure}

Our maps are comprised of C, D, and E array data and do not cover the
entire UV space. We do not have zero-spacing data which would be
required in order to reconstruct the most accurate map of the
region. The IRAM data consists of only a single pointing and its beam
does not cover the full area of the CARMA maps. In order to quantify
the amount of flux that could be resolved out by using CARMA, we
examined the single dish data taken by \citet{falgarone08}. Since the
IRAM-30 meter telescope has a resolution of 23\arcsec, we smoothed our
CARMA maps to match by convolving the data with a 23\arcsec\ Gaussian
beam (Figure \ref{iramcomp}). Spectra, (\ref{iramcom2}) were then
extracted from the position corresponding to the IRAM-30 meter
pointing center, as well as two other positions centered on the
north-eastern and north-western components. According to the IRAM
technical documentation \citep{iram30m}, the main beam efficiency is
78\% for the band that includes CN. To compare the flux values, an
estimate of CARMA's main beam efficiency is needed
\citep{white08}. CARMA does not have published main beam efficiency
measurements, however, it can be estimated from single-dish aperture
efficiency measurements which vary from 55\% to 70\% depending on the
individual dish. From this, we estimate that the main beam efficiency
(per dish) would be $\sim$70-80\%. By adopting a value of 75\%,
approximately the same as IRAM's, we can directly compare the IRAM and
CARMA spectra. The IRAM peak was 2.2 K, while CARMA's was 1.4 K. This
means that the peak of the CARMA spectra was at 64\% of IRAM's (the
area of the CARMA spectra is 48\% of the IRAM spectra). CN is also
seen in strong absorption towards the central source (W3OH), however,
this absorption feature is not present in any of the IRAM data. This
is possible for several reasons. First, the relatively small spatial
size of the absorption feature covers a small fraction of the total
IRAM beam. Any measured absorption would be diluted across the entire
beam and masked by the significantly greater emission. In addition,
the IRAM data baseline fitting procedure could contribute to hiding
the presence of a weak absorption feature.

\subsection{N$_2$H$^+$ Chemical Reaction}

In regions with standard CO abundances (i.e. regions that are not
depleted in CO), one of the formation mechanisms of HCO$^+$ is the
destruction of N$_2$H$^+$ by CO \citep{bergin02,jorgensen04}. Since
there is significant support that CN remains in the gas phase at
densities greater than 10$^5$ cm$^{-3}$ (same as N$_2$H$^+$) we would
expect CN and N$_2$H$^+$ to have a very similar spatial distribution,
however, there are some regions and velocities where we would expect
to see more N$_2$H$^+$ emission than we do. The particular reaction
that is thought to occur is:
\begin{equation}
N_2H^+ + CO \rightarrow HCO^+ + N_2
\end{equation}
If we compare the regions of strong N$_2$H$^+$ and HCO$^+$ emission,
we would expect them not to be coincident. In addition, we expect CO
to be depleted in regions of strong N$_2$H$^+$ emission, as the
presence of CO would cause the formation of HCO$^+$, while regions
with significant CO emission must be weak in N$_2$H$^+$, or the
reaction between these two molecules would occur.

In the region surrounding the Turner-Welch object we can see
considerable emission from both CO and HCO$^+$, however, there is no
N$_2$H$^+$ within our detection limit. This depletion of N$_2$H$^+$
could be caused by the amount of C$^{18}$O present along this
sightline. Since there is significant CN emission from this region, we
would also expect to see N$_2$H$^+$. The presence of CN and the
relative strengths of HCO$^+$ and C$^{18}$O is significant evidence of
this reaction occurring and that N$_2$H$^+$ is being consumed by this
reaction.

Above, we showed that there was a cold, dark region of gas towards the
continuum source where HCO$^+$ is seen in strong absorption along
with CN and HCN. Discounting the contribution from the continuum
source, C$^{18}$O is seen slightly in emission and N$_2$H$^+$ is seen
slightly in emission and absorption. The presence of CN absorption
implies conditions where N$_2$H$^+$ should also be strongly
detectable.  If we assume that the primary formation mechanism of
HCO$^+$ in this region is by the CO and N$_2$H$^+$ chemical reaction,
the relative depletion of these two chemical species in this cold dark
cloud can be understood.

In regions with conditions similar to that of W3(OH), CN may be a more
reliable tracer than N$_2$H$^+$ to directly sample regions at high
densities. However, since the strongest CN hyperfine transition has
significantly weaker line emission than the strongest hyperfine
transitions of N$_2$H$^+$, it would require significantly greater
telescope time to achieve a comparable signal-to-noise ratio. In our
case, these sets of maps show not only that this chemical reaction is
occurring, but is happening most strongly both along the sightlines
associated with an active star forming clump, and in a cold cloud in
front of an extremely bright UCHII region.

\subsection{Structure of the Region Surriounding W3(OH)}

\citet{wilson91} developed a model of the region surrounding W3(OH)
based on C$^{18}$O, C$^{34}$S, and methanol (among others). They
proposed a multi-layer model of low density molecular gas surrounding
a high density molecular core. This low density cooler region extends
in from of W3(OH) as well. Self-aborbtion in their CS data supports
this model. In addition, the $v_{LSR}$ of the self-absorbtion was up
to 2 km s$^{-1}$ more positive than the $v_{LSR}$ of the hotter
emitting region. This implys that the cloud envelope is contracting
relative to the cloud core. We see similar offsets in the E-array
spectrum of HCO$^{+}$ seen in Fig. \ref{hcospectra}a, however the
offset appears to decrease at higher angular resolutions
(Fig. \ref{hco+cuts}).

We can compare the range of radial velocities at which we see
molecular material. \citet{wilson91} data showed that C$^{18}$O peaks
towards W3(OH) at a radial velocity of -46.6 km s$^{-1}$, and peaks
towards the Turner-Welch object at -48 km s$^{-1}$. We see the same
radial velocities in our C$^{18}$O data (even though our dataset was
at 3mm, and theirs was at 1mm). For other species that trace related
density regimes (HCN and HCO$^{+}$), we see them peak at a velocity of
-47 km s$^{-1}$ towards the Turner-Welch object. These species are
seen strongly in absorbtion towards W3(OH), which makes it difficult
to determine its radial velocity. These data show that the region
around W3(OH) is extremely complex.

\section{Conclusions}

We mapped the 1 arcmin$^2$ region around W3(OH) with CARMA, in which
previous single-dish CN Zeeman observations detected a strong magnetic
field. In determining which gas was producing the magnetic field, we
were able to see the following features and make the following
conclusions from their presence:
\begin{enumerate}
\item A strong absorption feature in HCO$^+$, HCN, and CN was detected
  towards the continuum source whose strength and line saturation
  implies the existence of a cold, optically thick region in front of
  the continuum source.
\item Comparison between CARMA spectra and spectra obtained with the
  IRAM-30 meter telescope \citep{falgarone08} yields that CARMA
  detects approximately 65\% of the material in this region.
\item Selective depletion of N$_2$H$^+$ in comparison with the CN and
  HCO$^+$ distribution indicates that N$_2$H$^+$ is reacting with CO
  to form HCO$^+$. This reduces the effectiveness of N$_2$H$^+$ as a
  high-density tracer, and favors the use of species, such as CN, in
  objects with similar chemical composition to W3(OH). These
  measurments additionally support the usefulness of CN as a high
  density tracer and confirms the hypothesis that CN Zeeman mapping
  will probe the magnetic field strength in the high density regions
  of molecular cloud clumps.
\end{enumerate}
These conclusions lead to the result that future CN Zeeman mapping at
high resolution with an interferometer is feasible and that W3(OH) is
an ideal target to perform such mapping on.

\section{Acknowledgments}

Support for CARMA construction was derived from the states of
California, Illinois, and Maryland, the James S. McDonnell Foundation,
the Gordon and Betty Moore Foundation, the Kenneth T. and Eileen
L. Norris Foundation, the University of Chicago, the Associates of the
California Institute of Technology, and the National Science
Foundation. Ongoing CARMA development and operations are supported by
the National Science Foundation under a cooperative agreement (NSF AST
08-38226), and by the CARMA partner universities.

\begin{figure}
\begin{center}
\includegraphics[height=\linewidth, angle=-90]{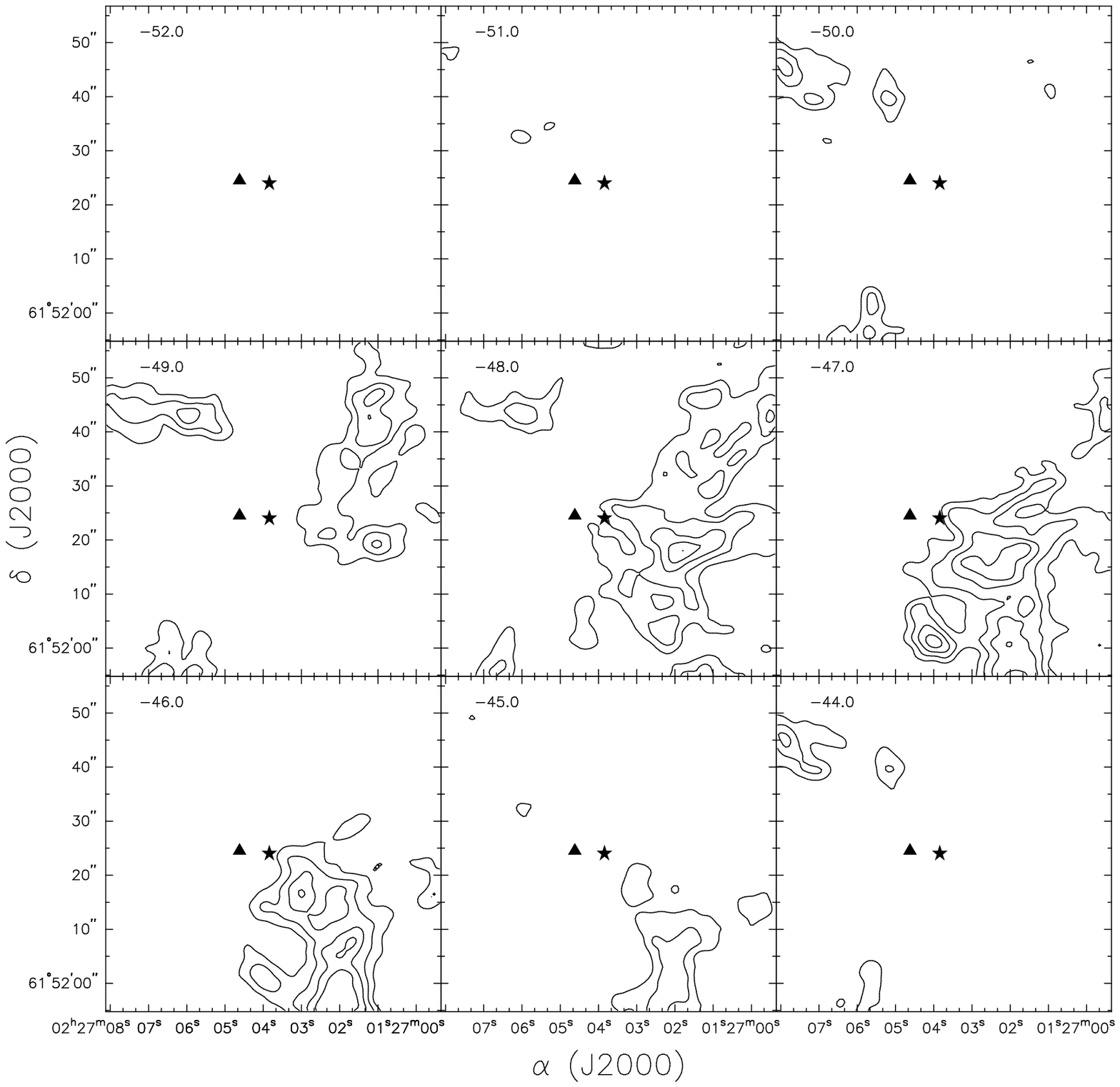}
\end{center}
\caption{N$_2$H$^+$ Channel maps, rebinned to 1 km s$^{-1}$ velocity
  increments. The velocity scale is centered on the strongest CN
  hyperfine component at 93.17348 GHz. The spacial resolution of this
  map is 3.4 x 2.6 arcsec. The contour levels are 20, 40, 60, 80, 100
  times 0.0129 Jy beam$^{-1}$. The $\bigstar$ and $\blacktriangle$
  represent the peak continuum positions of W3(OH) and the
  Turner-Welch object, respectively.
\label{n2h+channel}}
\end{figure}

\end{document}